\documentstyle[12pt]{article}
\begin{document}
\thispagestyle{empty}
\begin{center}
\LARGE \tt \bf {Irrotational vortex geometry of torsion loops}
\end{center}
\vspace{3.5cm}
\begin{center}
{\large By L.C. Garcia de Andrade\footnote{Departamento de F\'{\i}sica Te\'{o}rica - IF - UERJ - Rua S\~{a}o Francisco Xavier 524, Rio de Janeiro, RJ, Maracan\~{a}, CEP:20550.e-mail:garcia@dft.if.uerj.br}}
\end{center}
\begin{abstract}
The irrotational vortex geometry carachter of torsion loops is displayed by showing that torsion loops and nonradial flow acoustic metrics are conformally equivalent in $(1+1)$ dimensions while radial flow acoustic spacetime are conformally related in $(2+1)$ dimensional spacetime. The analysis of $2$-dimensional space allows us to express the fluid density in terms of the parameters of torsion loop metric. These results lead us to conclude that the acoustic metric of vortex flows is the gravitational analog of torsion loop spacetime. Since no vorticity in the fluids is considered we do not make explicit use of non-Riemannian geometry of vortex acoustics in classical fluids. Acoustic nonradial flows are shown to exihibit a full analogy with torsion loop metric.
\end{abstract}
\newpage

\section{Introduction}
A new method developed quite recently by M. Visser and Weinfurtner \cite{1}to show that the spacetime geometry of the equatorial slice of the Kerr black hole is formally equivalent to the geometry phonons follow in a rotating fluid vortex is used here to show that the phonon vortex geometry follow equivalent paths test particles would follow around a gravitational torsion loop \cite{2}. One of the main advantages of this approach is that this avoids the use of the recently introduced non-Riemannian geometry of vortex acoustics \cite{3} since here the curl of the speed of the fluid is given by a delta Dirac function and is only distinct from zero at the vortex site and not globally throughout the fluid. It is shown that in $(1+1)$ spacetime the nonradial flow of non-collapsing vortex geometry is equivalent to the teleparallel geometry of torsion loops. In the case of radial flows it is shown that $(2+1)$ dimensional vortex geometry fits quite well on the torsion loops geometry and is actually conformally mapped to this geometry. More recently Zhang et al \cite{4} have shown that acoustic wormholes can also be obtained following Visser-Weinfurtner method \cite{1}. The letter is organised as followed. Section 2 presents the vortex geometry of torsion loops in $(1+1)$ dimension, while in section 3 the radial flow vortex geometry in $(2+1)$ dimensions is conformally related to the torsion loop non-Riemannian geometry. In section 4 we present the acoustic nonradial flow and compute the coordinate transformation which makes the acoustic metric fully analog to the vortex geometry of acoustic flow. Also in this section by examining the $2$ dimensional vortex geometry one obtains a relation between the fluid speed and the torsion loop parameters as a step function. Since we are not dealing with rotational fluids we do not make use of the recently non-Riemannian geometry of rotational fluids \cite{3}. Since as we shall see the physical acoustic metric depends on the speed of the flow we could consider it as a Riemann-Finsler geometry \cite{5} with or without Cartan torsions.
\section{Vortex geometry of torsion loops in $(1+1)$ dimensions}
The physical acoustics metric \cite{1} is given by 
\begin{equation}
ds^{2}=\frac{{\rho}}{c}[-(c^{2}-{v_{\theta}}^{2})dt^{2}-2{v}_{\theta}rd{\theta}dt + dr^{2}+ r^{2}d{\theta}^{2}+ dz^{2}]
\label{1}
\end{equation}
and considering the cylindrically symmetric time-independent fluid flow endowed with a line vortex along the $z-axis$, the flow velocity $\vec{v}$ reads  
\begin{equation}
\vec{v}= v_{r}(r){\vec{e}}_{r}+{v}_{\theta}{\vec{e}}_{\theta}
\label{2}
\end{equation}
where $\vec{e}_{r}$ and $\vec{e}_{\theta}$ are respectively the unit base vectors of the planar flow. The conservation equation and Euler equation for the irrotational flow are written as 
\begin{equation}
{\nabla}.\vec{v}=0
\label{3}
\end{equation}
\begin{equation}
\vec{f}={\rho}(\vec{v}.{\nabla}){\vec{v}}+c^{2}{\partial}{\rho}{\vec{e}}_{r}
\label{4}
\end{equation}
A simple analysis of the torsion loop metric in $(1+1)$ dimension
\begin{equation}
{ds^{2}}_{{TL}_{(1+1)}}= -dt^{2}-2{B}_{\theta}adtd{\theta}+(1-{B^{2}}_{\theta})a^{2}d{\theta}^{2} 
\label{5}
\end{equation}
in $(t,{\theta})$ allows us to infer the conformal relation between the vortex geometry of non-radial flow, where $v_{r}=0$, and the the torsion loop metric
\begin{equation}
{ds^{2}}_{(1+1)}={\rho}[1-{v_{\theta}}^{2}]{ds^{2}}_{{TL}_{(1+1)}}
\label{6}
\end{equation}
where $c^{2}(r=a)=1$ where $a$ is the radius of the torsion loop. An important feature of this geometry is that the flow goes supersonic ${v_{\theta}}^{2}= c^{2}=1$ and due to the conformal relation between the acoustic and torsion loop metrics implies $ds^{2}=0$. 
\section{Vortex geometry of torsion loops in $(2+1)-dimensional$ spacetime}
Let us now consider the acoustic metric in $(2+1)$ dimensional effective spacetime for the non-radial flow
\begin{equation}
{ds^{2}}_{(2+1)}= \frac{\rho}{c}(c^{2}-{v_{r}}^{2}-{v_{\theta}}^{2})[-dt^{2}-\frac{(2{v}_{\theta}rdtd{\theta}+2{v}_{r}dtdr-dr^{2}-r^{2}d{\theta}^{2})}{(c^{2}-{v_{r}}^{2}-{v_{\theta}}^{2})}]  
\label{7}
\end{equation}
which along with the torsion loop metric in $(2+1)$ spacetime 
\begin{equation}
{ds^{2}}_{{TL}_{(2+1)}}= -dt^{2}-2{B}_{\theta}rdtd{\theta}+(1-{B^{2}}_{\theta})r^{2}d{\theta}^{2}-2{B}_{r}dtdr+(1-{B^{2}}_{r})dr^{2}
\label{8}
\end{equation}
yields the following constraint equations
\begin{equation}
B_{\theta}=\frac{v_{\theta}}{c}
\label{9}
\end{equation}
\begin{equation}
B_{r}=\frac{v_{r}}{c}
\label{10}
\end{equation}
where we adopt the approximation $({v_{r}}^{2}+{v_{\theta}}^{2})<<c^{2}$ which is a subsonic approximation since $c(r)$ represents the speed of sound in the fluid. This time the conformal equivalence between the metrics is
\begin{equation}
{ds^{2}}_{(2+1)}={\rho}{ds^{2}}_{{TL}_{(2+1)}}
\label{11}
\end{equation}
By considering the $2-dimensional$ acoustic and torsion loop metrics
\begin{equation}
ds^{2}=\frac{{\rho}}{c}[dr^{2}+r^{2}d{\theta}^{2}]
\label{12}
\end{equation}
and
\begin{equation}
ds^{2}_{TL_{2}}= (1-{B_{\theta}}^{2})[dr^{2}+r^{2}d{\theta}^{2}]
\label{13}
\end{equation}
allows us to infer the following relation between the fluid density and the metric coefficient $B_{\theta}$
\begin{equation}
\frac{{\rho}(r)}{c(r)}=(1-{B^{2}}_{\theta})
\label{14}
\end{equation}
to carry out the computations in this section we consider the degenerate case where $B_{r}=B_{\theta}$. Finally we mention that due to the equivalence shown here one may rewrite the Letelier constraint \cite{2} 
\begin{equation}
{\nabla}{\times}{\vec{B}}={\vec{J}}
\label{15}
\end{equation}
where $\vec{J}$ is the Cartan torsion vector, as
\cite{2} 
\begin{equation}
{\nabla}{\times}\frac{\vec{v}}{c^{2}}={\vec{J}}=J_{0}{\delta}(r-a){\delta}(z)
\label{16}
\end{equation}
where ${\delta}(x)$ represents the Dirac delta distribution and $\vec{J}$ in this case is the torsion vector of the vortex geometry along the vortex z-direction. 
\section{Acoustic nonradial flows}
To show that Letelier torsion loop metric is analogous to the vortex bath tub geometry we consider the following form of the nonradial $(B_{r}=0)$ form of torsion loop metric
\begin{equation}
{ds^{2}}_{{TL}_{(2+1)}}= -dt^{2}-2v_{\theta}rdtd{\theta}+dr^{2}+(1-{B^{2}}_{\theta})r^{2}d{\theta}^{2}
\label{17}
\end{equation}
Comparison of this metric with metric 
\begin{equation}
{ds^{2}}_{(2)}= {\Omega}^{2}(r')(d{r'}^{2}+{r'}^{2}d{\theta}^{2})
\label{18}
\end{equation}
yields the following equations 
\begin{equation}
{\Omega}^{2}(r'){r'}^{2}=(1-B_{\theta})r^{2}
\label{19}
\end{equation}
\begin{equation}
{\Omega}^{2}(r')d{r'}^{2}= dr^{2}
\label{20}
\end{equation}
From this couple system of equations one obtains
\begin{equation}
\frac{dr^{2}}{r^{2}}=\frac{(1-B_{\theta})}{{r'}^{2}}d{r'}^{2}
\label{21}
\end{equation}
A simple integration can be obtained by considering the expansion $(1-{B_{\theta}}^{2})^{\frac{1}{2}}=(1+\frac{1}{2}{B^{2}}_{\theta})$ which yields 
\begin{equation}
r'= {r}^{\frac{{J_{0}}^{2}}{2}}+c
\label{22}
\end{equation}
where c is an integration constant. During the integration we have used that 
\begin{equation}
B_{\theta}=J_{0}H_{0}(r-a)
\label{22}
\end{equation}
where $H_{0}(r-a)$ is the Heavised step function which is $1$ when the $r$ is greater or equal the torsion loop radius $a$. Expression (\ref{22}) represents the solution of the vortex equation 
\begin{equation}
{\epsilon}_{zr{\theta}}{\partial}_{r}B_{\theta}= J_{0}{\delta}(r-a)
\label{23}
\end{equation}
This is trivial since the derivative of the Heaviside function is the Dirac delta distribution. Comparison between metrics (\ref{7}) and 
\begin{equation}
{ds^{2}}_{{TL}_{(2+1)}}= {\Omega}^{2}[{\Omega}^{-2}(-dt^{2}-2v_{\theta}rdtd{\theta})+(d{r'}^{2}+(1-{B^{2}}_{\theta})r^{2}d{\theta}^{2}]
\label{24}
\end{equation}
yields the following expressions for the analog model
\begin{equation}
\frac{{\rho}}{c}={\Omega}^{2}
\label{25}
\end{equation}
\begin{equation}
v_{\theta}={\Omega}^{-2}J_{0}H_{0}(r-a)
\label{26}
\end{equation}
which completely carachterizes the "flow". Expression (\ref{25}) was also obtained for the vortex geometry of the equatorial plane of the black hole in reference \cite{4}. Finally one might notice that the torsion loop behaves analously to the bath tub since the step function tells us that for $r<a$ or inside the loop, or in the analog model inside the bath tub , the velocity vanishes, or there is no fluid in this region since in the plane of the bath tub there the flow has only z-component and no component on the ${\theta}$ direction. 

\section{Discussion and conclusions}
We have shown that vortex Finsler type effective geometry can be conformally mapped into the teleparallel geometry of a torsion loop metric as long as some constraints on fluid flow and dimension be respected. From the discussion and formulas presented here one could easily compute the Euler equations for the equivalent gravitational analog. We finally show the analogy between the gravitational torsion loop and the bath tub metric. 
\section*{Acknowledgements}
 I would like to express my gratitude to P.S. Letelier, C. Furtado and S. Bergliaffa for helpful discussions on the subject of this paper. Grants from CNPq. and UERJ are grateful acknowledged.

\end{document}